# Cholic acid-based mixed micelles as siRNA delivery agents for gene therapy

*Alexander J. Cunningham[1,2], Victor Passos Gibson[2,3], Xavier Banquy[2], X.X. Zhu[1], and Jeanne Leblond Chain[2,3]\**

[1] Département de Chimie, Université de Montréal, C.P. 6128, Succ. Centre-ville, Montréal, QC, H3C 3J7, Canada

[2] Faculté de Pharmacie, Université de Montréal, C.P. 6128, Succ. Centre-ville, Montréal, QC, H3C 3J7, Canada

[3] Gene Delivery Laboratory, Faculté de pharmacie, Université de Montréal, H3C 3J7, Montréal, QC, Canada

\* Correspondence to: jeanne.leblond-chain@inserm.fr. Present address: ARNA Laboratory, Faculty of pharmacy, Université Bordeaux, INSERM U1212, CNRS UMR 5320, F-33016 Bordeaux, France.

# Abstract

Gene therapy is a promising tool for the treatment of various cancers but is hindered by the physico-chemical properties of siRNA and needs a suitable vector for the delivery of siRNA to the target tissue. Bile acid-based block copolymers offers certain advantages for the loading and delivery of siRNA since they can efficiently complex siRNA and bile acids are biocompatible endogenous molecules. In this study, we demonstrate the use of lipids as co-surfactants for the preparation of mixed micelles to improve the siRNA delivery of cholic acid-based block copolymers. Poly(allyl glycidyl ether) (PAGE) and poly(ethylene glycol) (PEG) were polymerized on the surface of cholic acid to afford a star-shaped block copolymer with four arms (CA-PAGE-*b*-PEG)$_4$. The allyl groups of PAGE were functionalized to bear primary or tertiary amines and folic acid was grafted onto the PEG chain end to increase cell uptake. (CA-PAGE-*b*-PEG)$_4$ functionalized with either primary or tertiary amines show high siRNA complexation with close to 100% complexation at N/P ratio of 8. Uniform aggregates with diameters between 181 and 188 nm were obtained. DOPE, DSPE-PEG$_{2k}$, and DSPE-PEG$_{5k}$ lipids were added as co-surfactants to help stabilize the nanoparticles in the cell culture media. Mixed micelles had high siRNA loading with close to 100% functionalization at N/P ratio of 16 and diameters ranging from 153 to 221 nm. The presence of lipids in the mixed micelles improved cell uptake with a concomitant siRNA transfection in HeLa and HeLa-GFP model cells, respectively.

# Key words

siRNA delivery, polymer, mixed micelle, cholic acid, poly(ethylene glycol), lower critical solution temperature

# 1. Introduction

In recent years, a great deal of effort has been directed towards the development and optimization of nanoparticles for gene delivery. Gene delivery offers a formulation challenge due to the hydrophilicity, large size, and numerous negative charges on the siRNA. Nevertheless, a vast array of publications describing gene delivery systems has flooded the scientific literature with many reaching clinical trials but very few currently used in the clinics (Kaczmarek et al., 2017). The two most thoroughly studied non-viral systems are lipid- and polymer-based nanoparticles, both characterized by their own limitations. Polymers have the advantage of low immunogenicity, ease of synthesis with a lower production cost, and versatility (Lai and Wong, 2018; Oupicky, 2016; Xiang et al., 2017), but they are mainly limited by their cytotoxicity and low transfection potential. The cytotoxicity arises from the high amount of positive charges needed to complex the gene product, whereas the low transfection potential is a result of low endosomal escape once internalized inside the cell (Degors et al., 2019; Zhou et al., 2017). Polyethyleneimine (PEI) has been extensively studied for its high transfection potential (Lai and Wong, 2018). There exists a balance in PEI between high transfection and good biocompatibility governed by the length of the polymer that is directly linked to the amount of positive charges, where 25 kDa is considered the gold standard (Huh et al., 2007; Zhou et al., 2017). Interestingly, reports have demonstrated that branched PEI was more efficient than linear PEI in complexing siRNA with lower N/P ratios needed to complex the siRNA (Boeckle et al., 2004; Scholz and Wagner, 2012). It is believed that the branched structure of PEI allows more flexibility and folding options in a 3-dimensional space to complex the siRNA that is thought to be more rigid and less amenable to folding and rearrangement (Scholz and Wagner, 2012). Cholic acid-modified PEI have been shown to increase the transfection potential of low molecular weight PEI (Dube et al., 2018; Kim et al., 2012; Weecharangsan et al., 2017). Indeed, adding amphiphilic bile acids drove the formation of polyplexes of low molecular weigth PEI with siRNA and improved gene delivery. Furthermore, different reports have demonstrated the advantage of using branched polymers. For example, a branched histidine/lysine peptide was shown to effectively inhibit protein expression by 80% (Zhou et al., 2016). Dendrimers, hyperbranched structures, are also gaining importance in the field of siRNA delivery (Dzmitruk et al., 2018; Yang et al., 2015). Their branched structure allows a high loading of siRNA (Dzmitruk et al., 2018).

To further improve the transfection efficiency of polymer-based siRNA delivery systems while minimizing the cytotoxic effects, we demonstrate the use of cholic acid-based star-shaped block copolymers. Cholic acid (CA) is an endogenous compound serving in the emulsification of fats and lipids for their uptake throught the intestinal cell wall. Moreover, cholic acid has four hydroxyl groups that can serve as initiators for polymerization yielding a star shaped structure. Recent reports have demonstrated its potential as a drug delivery system due to its inherent capacity to form micelles and its low cytotoxicity (Cunningham and Zhu, 2016). PEG-functionalized CA has been successfully used in the loading of itraconazole showing high loading content (35 %) (Le Dévédec et al., 2013). In another report, CA-based diblock copolymers showed high loading content (14 %) of Doxorubicin via electrostatic interaction with minimal cytotoxicity (Cunningham et al., 2018). In this report, CA was grafted with four chains of poly(allyl glycidyl ether) (PAGE) and poly(ethylene glycol) (PEG) blocks via anionic polymerization to form a star-shaped block copolymer with 4-arms, CA-(PAGE-*b*-PEG)$_4$. The allyl groups of PAGE were functionalized to bear amine groups promoting the siRNA complexation. Based on the studies of branched PEI, the star-shaped architecture is believed to promote a higher complexation with siRNA due to the arrangement of the arms around the rigid siRNA molecules. To further stabilize the CA-based block copolymer formulations and promote a higher cell uptake, lipids like DOPE, DSPE-PEG$_{2k}$ and DSPE-PEG$_{5k}$ were added to form mixed micelles. Adding lipids to polymers in mixed micelles is proposed to help in the stability of the carrier system and loading of active pharmaceutical ingredients (Harmon et al., 2011; Le Devedec et al., 2013; Li et al., 2011b; Wu et al., 2013). DOPE, a neutral lipid, is known to promote endosomal escape (Hoekstra et al., 2007), whereas DSPE-PEG has been shown to have good stabilization potential and is currently used for the formulation of the FDA-approved Doxil® (Li et al., 2015). In this report, the potential of the CA-based block copolymers toward siRNA delivery is measured in the absence and presence of the lipids as co-surfactants and the transfection is compared.

## 2. Materials and Methods
## 2.1 Materials

All chemical reagents were purchased from Sigma-Aldrich and used without further purification unless stated otherwise. For the anionic polymerization, dimethyl sulfoxide (DMSO) and allyl glycidyl ether (AGE) were dried overnight with calcium hydride and distilled immediately prior to use. Ethylene oxide in the gaseous state was passed through a column of calcium hydride and condensed with dry ice and acetone for quantification by volume before adding to the reaction vessel. Tetrahydrofuran (THF) was dried with sodium under reflux, and methanol was dried using magnesium sulfate. siRNA GFP duplex (target sequence 5'-GCA AGC TGA CCC TGA AGT TC–3') was purchased from Dhermacon (Lafayette, CO, United States) and negative control was purchased from Alpha DNA (Montreal, QC, Canada) Sense strand: UAGCGACUAAACACAUCAAUU and antisense strand: UUGAUGUGUUUAGUCGUAAUU. The fluorescently-tagged siRNA used for quantifying cell uptake with flow cytometry was the negative siRNA tagged with Alexa Fluor 488 purchased from Qiagen (Toronto, ON, Canada). The siRNA used for cell uptake was tagged with Cy5 and was obtained from Dhermacon (Lafayette, CO, United States) Sense Strand: 5'-Cy5-UAGCGACUAAACACAUCAAUU-3' Antisense strand: 5'-UUGAUGUGUUUAGUCGCUAUU-3' with the 5'-Cy5 modification on the sense strand.

## 2.2 Synthesis of CA-(PAGE-*b*-PEG)$_4$ via anionic polymerization

The synthesis and functionalization of the polymers are depicted in Scheme 1. The ethanolamine derivative of cholic acid with four hydroxyl groups was synthesized according to a previous method (Gouin and Zhu, 1996; Luo et al., 2009). For anionic polymerization, the glassware was flame-dried under vacuum and purged three times with argon before use. Briefly, 5β-cholanoamide (1.1 mmol, 0.5 g) was dissolved in 40 mL of dry DMSO. A potassium naphthalenide solution in THF was added (0.4 mol/L, 1.1 mmol, 1 eq., 3.2 mL) dropwise using a canula under high pressure. AGE was distilled and added (5.1 g, 44.3 mmol, 40 eq.) dropwise using a canula under high pressure. The anionic polymerization was initiated by immersing in an oil bath at 40 °C and allowed to proceed for 24 h to allow the consumption of all the AGE monomers. Then, dry ethylene oxide (6.6 mL, 132.9 mmol, 120 eq.) was chilled in dry ice/acetone to condense into a liquid for accurate measuring and was introduced into the flask. The mixture was polymerized for 24 h. The reaction was quenched by addition of concentrated hydrochloric acid (1 mL). The DMSO solution was extracted with hexane (3 x 10 mL) to remove the naphthalenide. Distilled water was added to the DMSO solution and the mixture was dialyzed against distilled water (48 h) through a regenerated cellulose membrane (MWCO 3,500

Da) (Fisher Scientific, Ottawa, ON, Canada) to remove all unreacted monomers. The polymer was then lyophilized and stored at 4 °C.

## 2.3 Functionalization of CA-(PAGE-*b*-PEG)$_4$ by addition of amine groups

For the preparation of CA-(PAGE-NH$_2$-*b*-PEG)$_4$ (or ABP-NH$_2$) containing primary amine groups, the purified and dried polymers of CA-(PAGE-*b*-PEG)$_4$ (2 g, 0.3 mmol) were dissolved in dry methanol (MeOH) (25 mL). 2,2′-Azobis(2-methylpropionitrile) (AIBN) (1.6 g, 9.7 mmol) was dissolved in MeOH. Cysteamine hydrochloride (8.8 g, 77.6 mmol) was dissolved in MeOH and the reaction was started by immersion in an oil bath at 75 ºC under reflux and Ar atmosphere. The reaction proceeded for 12 h, was cooled to room temperature and purified by dialysis against distilled water (48 h, MWCO 3,500 Da) to remove all unreacted cysteamine. A similar protocol was followed for the synthesis of CA-(PAGE-N(Et)$_2$-*b*-PEG)$_4$ (or ABP-NEt$_2$), however THF was used rather MeOH due to the limited solubility of 2-diethylaminoethanethiol HCl in MeOH. CA-(PAGE-*b*-PEG)$_4$ (1 g, 0.08 mmol) was dissolved in dry THF. AIBN (0.4 g, 2.2 mmol) and 2-Diethylaminoethanethiol HCl (3.8 g, 22.4 mmol) were dissolved in THF. The reaction was started by immersion in an oil bath at 75 ºC under reflux and Ar atmosphere. The reaction proceeded for 12 h and was cooled to room temperature. The reaction was purified by dialysis against distilled water (48 h, MWCO 3,500 Da) to remove all unreacted 2-diethylaminoethanethiol. The polymer was then lyophilized and stored at 4 °C.

## 2.4 Folic acid conjugation to CA-(PAGE-NEt$_2$-*b*-PEG)$_4$

In a round bottom flask, CA-(PAGE-NEt$_2$-*b*-PEG)$_4$ (1 g, 0.07 mmol) was dissolved in 15 mL dry DMF. Folic acid (FA) (7.8 mg, 0.07 mmol), N,N′-dicyclohexylcarbodiimide (DCC) (14 mg, 0.07 mmol), and 4-(dimethylamino)pyridine (DMAP) (8 mg, 0.07 mmol) were dissolved in DMF. The mixture was stirred in an oil bath at 30 ºC for 48 h. Then, the reaction was cooled to room temperature and filtered. The product was purified by dialysis against distilled water (48 h, MWCO 3,500) to remove all unreacted FA. The polymer (ABP-NEt$_2$-FA) was then lyophilized and stored at 4 °C.

## 2.5 Characterization methods

The molar mass of the polymers was determined by size exclusion chromatography (SEC) running on DMF as eluent on a Breeze system from Waters equipped with a 717 plus autosampler, a 1525 Binary HPLC pump, and a 2410 refractive index detector and two consecutive Polar Gel M and Polar Gel L columns (Agilent technologies, Santa Clara, CA, United States) with a flow rate of 0.7 mL/min and at 50°C. Polymethyl methacrylate (PMMA) standards were used for calibration. All samples were filtered on a nylon 0.2 μm filters prior to injection. $^1$H-NMR spectra were recorded on a Bruker AV400 spectrometer operating at 400 MHz and samples were dissolved in $d_6$-DMSO. Dynamic light scattering (DLS) measurements were performed on a Malvern Zetasizer NanoZS instrument equipped with a He-Ne laser with a wavelength of 633 nm, and at a scattering angle of 173.5°. Intensity-averaged hydrodynamic diameters of the dispersions were obtained using the non-negative least-squares algorithm (NNLS). Disposable micro-cuvettes were used, and the samples (0.5 mg/mL) were filtered using 0.45 μm nylon filters prior to measurements. Samples were run at 37 ºC in the appropriate solvent (5 % Dextrose or DMEM). When using DMEM cell culture media, the parameters for running the DLS measurements were obtained from a previous publication: viscosity 0.94 cP, refractive index 1.345, dielectric constant 77.5896 (Semisch et al., 2014). Lower critical solution temperatures (LCST) of the polymer samples (1 mg/mL) were analyzed by following the temperature trend of the polymer with DLS. The sequence was set at a starting temperature of 20 ºC and an end temperature of 60 ºC with a 1 ºC temperature interval. The sample equilibrated at each temperature interval for 120 sec and 3 measurements were obtained for each interval. The general-purpose analysis model was used. The zetapotential was obtained on the same Malvern Zetasizer NanoZS instrument. The samples (0.03 mg/mL in distilled water) were run at 37 ºC using the Smoluchowski approximation and monomodal analysis model. Differential scanning calorimetry (DSC) was performed on a DSC Q1000 (TA instruments) at a heating rate of 10 °C/min. The temperature and heat flow were calibrated with indium before each measurements.

## 2.6 siRNA encapsulation

A stock solution of siRNA scrambled (AlphaDNA, Montreal, QC, Canada) was prepared in 5% Dextrose (Dex) (160 nM). Stock solutions of polymer were prepared in 5% Dex at appropriate conditions to afford nitrogen of polymer /phosphate of siRNA (N/P) ratios ranging from 1 to 32 (using 42 phosphate groups for siRNA). 250 μL of siRNA solution was mixed with 250 μL of polymer solution and briefly vortexed. The complex was incubated for 30 min at room temperature. Then, an aliquot of 100 μL was added to a 96 well plate and 40 μL of 4X SYBR Gold fluorescent dye (Thermo Scientific, Waltham, MA, United States) was added to the aliquot and the fluorescence was measured using a Safire microplate reader (Tecan, Seestrasse, Switzerland) ($\lambda_{ex}$ = 495 nm, $\lambda_{em}$ = 537 nm). Three aliquots of each N/P ratio were measured as triplicates. To quantify the amount of siRNA loaded, a calibration curve was constructed. siRNA with concentrations ranging from 0 to 120 nM were added to a plate reader (100 μL) and 40 μL of 4X SYBR Gold was added to these solutions. The fluorescence of the SYBR Gold was measured and plotted into a calibration curve. The siRNA encapsulation efficiency (EE) was calculated from the siRNA concentrations using the following formula:

$$EE = \frac{[siRNA]\ in\ feed - [siRNA]\ measured\ with\ SYBR\ Gold}{[siRNA]\ in\ feed} * 100\ \%$$

## 2.7 Cytotoxicity

HeLa-GFP cells (Cell Biolabs Inc, San Diego, CA USA) and HeLa cells (CCL-2TM, ATCC®) were cultured in Dulbecco's Minimum Essential Medium (DMEM) supplemented with 10% fetal bovine serum (FBS) (Gibco, Burlington, ON, Canada). The cells were maintained in an incubator at 37 ºC and a 5% $CO_2$ atmosphere. MTS assay (Sigma Aldrich, Oakville, ON, Canada) was used to determine the cytotoxicity of the blank formulations and the siRNA-loaded formulations. MTS assay (Abcam ab197010, Toronto, ON, Canada) was conducted according to manufacturer's protocol. HeLa cells were plated on a 96-well plate at a seeding density of 5,000 cells per well in 200 μL DMEM and allowed to adhere overnight (37°C, 5% $CO_2$, 12 h). 40 μL of the formulation to be tested was added to each well. The plates were incubated for 48 h (37°C, 5% $CO_2$). 10 μL of MTS solution was added to each well and incubated (37°C, 5% $CO_2$) for 30, 60 and 120 min, respectively. The absorbance was read at 490 nm using a plate reader (Tecan, Seestrasse, Switzerland) at each time point. Cytotoxicity is reported as compared to absorbance measured for control (untreated cells).

## 2.8 Transfection of siRNA-loaded nanoparticles

To assess the silencing capabilities of the siRNA-loaded nanoparticles, a HeLa-GFP cell line was used. HeLa-GFP cells (Cell Biolabs Inc, San Diego, CA USA) were cultured in Dulbecco's Minimum Essential Medium (DMEM) supplemented with 10% fetal bovine serum (FBS) (Gibco, Burlington, ON, Canada). The cells were maintained in an incubator at 37 ºC and a 5% $CO_2$ atmosphere. HeLa-GFP cells were placed in a 12-well plate at a seeding density of 40,000 cells/well in 1 mL DMEM/FBS and allowed to adhere overnight (37°C, 5% $CO_2$). The following day, the cells were rinsed with PBS and 1 mL of Opti-MEM was added to the wells. The siRNA-loaded formulations were prepared in 5% Dex as described above using the corresponding N/P for the polymer being assayed and either 20, 40 or 60 nM siRNA. The complex was vortexed and incubated for 30 min. Then, the formulations were added to the cells, centrifuged (800 rpm, 5 min) and incubated for 4 h. After the incubation period, the cells were washed with PBS and fresh DMEM/FBS (1 mL) was added. The cells were incubated for 48 h. After this incubation period, the cells were washed with PBS, trypsinized, and suspended in FACS buffer (95 % 1X PBS, 5 % FBS, 0.1% sodium azide) for analysis using a FACSCalibur® flow cytometer (BD Biosciences, San Jose, CA, USA). GFP expression was measured for each triplicate sample using FlowJo software and taking the mean fluorescence intensity of the cell population (minimum 10 000 events). The GFP expression was calculated taking untreated samples as control samples and using this control as a relative measure to calculate the % GFP expression in the siRNA treated samples. A positive control included Lipofectamine RNAiMAX® (Thermo Scientific) according to the manufacturer's protocol and a negative control included HeLa cells not expressing the GFP protein. All experiments were conducted in triplicates.

## 2.9 Cell uptake

HeLa cells were grown in DMEM medium supplemented with 10% FBS. Cells were seeded in a 12-well plate at a seeding density of 100,000 cells/well and allowed to adhere overnight (37°C, 5% $CO_2$). The following day, the cells were rinsed with PBS and DMEM/FBS was replaced with Opti-MEM (1 mL). The formulations were prepared as above, but using siRNA labeled with Alexa488 fluorophore (60 nM, Qiagen, Toronto, ON, Canada) with the appropriate N/P ratio. After the corresponding incubation time (2, 4, and 24 h), the cells were washed with PBS, trypsinized (100 μL of 0.25 % Trypsin/EDTA), and suspended in FACS buffer (400 μL, 95 % PBS, 5 % FBS, 1 mM EDTA). The cells were observed on a FACScalibur flow cytometer (BD Biosciences, San Jose, CA, USA). Untreated HeLa cells were used as negative control and the % cell uptake were calculated using Lipofectamine RNAiMAX® (Thermo Scientific) as positive control. The experiment was done in triplicate.

For the fluorescence microscopy, the same procedure was followed with the exception of a fixed 4 h incubation time for the siRNA-labeled formulations and a siRNA labeled with Cy5.5 (Dhermacon, Lafeyette, CO, United States). Alternatively, 100 nM of LysoTracker® Green (Thermo Scientific) was added 30 min prior to imaging for the colocalization of siRNA and lysosomes. Imaging of HeLa cells was performed using an Olympus IX81 fluorescent microscope equipped with a Plan Apo N 40X silicone objective (Olympus Canada Inc., Toronto, ON, Canada) and a 12 bits Retiga-2000R CCD camera (QImaging, Surry, BC, Canada). The images were obtained and analyzed using MetaMorph Advanced software 7.8.9 (Molecular Devices, San Jose, CA, USA). The Cy5 channel ($\lambda_{ex}$ = 649 nm, $\lambda_{em}$ = 666 nm) and the green channel ($\lambda_{ex}$ = 504 nm, $\lambda_{em}$ = 666 nm) were used. All fluorescent images were exported keeping the scaling and grey level range constant. The experiment was done in triplicate.

## 3. Results
### 3.1 Polymer synthesis and characterization

Star-shaped block copolymers bearing pendant amine groups were prepared via anionic polymerization followed by post-synthesis modifications according to Scheme 1. An ethanolamine derivative of cholic acid was obtained via a condensation reaction according to a previously published protocol (Gouin and Zhu, 1996; Luo et al., 2009). This cholic acid derivative was used as an initiator in the anionic polymerization of allyl glycidyl ether (AGE), followed by the polymerization of ethylene oxide (EG) to afford a four-branched block copolymer CA-(PAGE-*b*-PEG)$_4$.

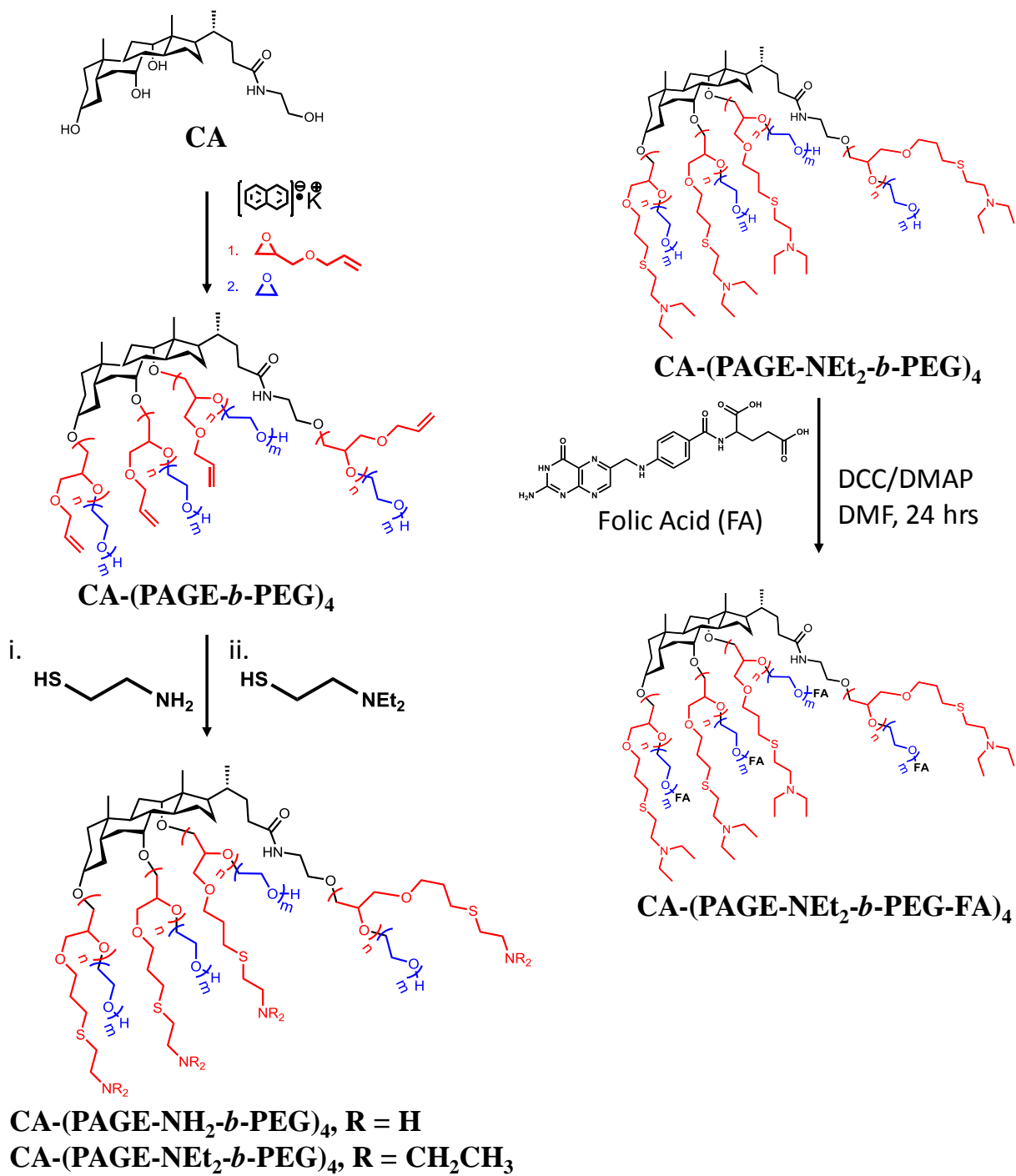

**Scheme 1.** Synthesis and functionalization of cholic acid-based nanoparticles.

The copolymers obtained were analyzed using SEC and $^1$HNMR spectroscopy (Table 1). Two polymers present similar length of PAGE and variable PEG length to study the effect of the latter on the siRNA encapsulation efficiency. The SEC traces show an increase in the $M_n$ when the PEG block was added on CA-(PAGE)$_4$ to form CA-(PAGE-$b$-PEG)$_4$ and a monomodal distribution with a low dispersity index (Figure S1). In the $^1$HNMR spectrum (Figure S2), the molar ratio of both polymers was calculated based on the methine peak (5.79-5.92 ppm) of PAGE and methylene peak (3.35-3.60 ppm) of PAGE and PEG, using the methyl proton (0.58 ppm) of cholic acid skeleton as reference. According to the $^1$HNMR integration, the two polymers possess the following architecture: CA-(AGE$_6$-*b*-EG$_{17}$)$_4$ and CA-(AGE$_8$-*b*-EG$_{49}$)$_4$ (Table 1). PAGE blocks were further functionalized with primary and tertiary amine groups through thiol-ene reactions, in order to promote electrostatic interactions with siRNA, and study their potential towards transfection while minimizing the cytotoxicity of the polymers. It was previously reported that primary amines show permanent positive charge under various pH conditions and this high positive charge leads to stronger binding and hence better siRNA complexation, but can also give rise to inefficient release of the siRNA in the cytosol along with a high cytotoxicity (Knudsen et al., 2015; Lin et al., 2019; Obata et al., 2008). Teritiary amines show a lower surface potential, therefore a lower complexation strength which can be beneficial for cytosolic release and lower cytotoxicity (Lu et al., 2016; Whitehead et al., 2014). Tertiary amines have a lower pK$_a$ and can protonate in the endosome to help in endosomal escape (Degors et al., 2019; Habrant et al., 2016; Mo et al., 2013). The purified polymers were therefore reacted with a thiol-containing aliphatic amine (primary and tertiary amine) via thiol-ene reaction (Scheme 1) and the complete functionalization was confirmed by $^1$HNMR (Figure S3 and S4). Upon functionalization, the allyl peaks (5.79-5.92 ppm) disappear and new peaks appear (2.5-3 ppm) corresponding to the 2-(diethylamino)ethanethiol moiety. Using the reference peak at 0.58 ppm corresponding to the methyl protons on the CA steroid skeleton, $^1$HNMR results show close to 100% functionalization for the coupling of diethylaminomethyl and cysteamine. Finally, to improve the cell uptake of the siRNA-loaded NP, folic acid (FA) was coupled as an active targeting ligand to the terminal hydroxyl groups of the PEG chains of the tertiary-amine functionalized block copolymer. Folate receptors are over-expressed on the surface of various cancer cell lines whereas in healthy cells they are restricted to cells of the lung, kidneys, placenta, and choroid plexus with lower expressions (Parker et al., 2005; Yoo and Park, 2004; Zwicke et al., 2012). Therefore, as demonstrated in the literature, coupling FA to the surface of NP has the potential to increase their uptake in cancer cells over healthy cells (Yoo and Park, 2004; Zwicke et al., 2012). FA was coupled using a Steglich esterification following a previously published procedure (Cao et al., 2018). The purified polymers were analyzed by $^1$HNMR and UV-Vis spectroscopy. $^1$HNMR (Figure S5) shows the appearance of characteristic aromatic peaks for FA between 6.6-8.2 ppm. The functionalization ratio was calculated based on the FA methine peaks (region 7.7 ppm) as compared to the cholic acid reference peak (0.58 ppm). The results indicate that 62 % of the polymers were functionalized. Although UV-Vis spectroscopy cannot be used to quantify the amount of FA present on the polymers due to differences in the molar absorption coefficient of free FA and polymer-bound FA (data not shown), the UV-Vis spectrum corroborates the successful presence of FA on purified polymers (Figure S6).

**Table 1. $^1$H-NMR and SEC characterization of CA-(PAGE-*b*-PEG)$_4$ polymers.**

|  | Molecular weight (g/mol) | | |
| --- | --- | --- | --- |
| Polymer | $^1$HNMR | SEC | Đ |
| CA-(AGE$_6$)$_4$ | 3,000 | 3,000 | 1.09 |
| CA-(AGE$_6$-*b*-EG$_{17}$)$_4$ | 6,200 | 5,000 | 1.13 |
| CA-(AGE$_8$)$_4$ | 4,300 | 5,500 | 1.19 |
| CA-(AGE$_8$-*b*-EG$_{49}$)$_4$ | 13,000 | 16,200 | 1.18 |

Đ: polydispersity

### 3.2 siRNA encapsulation

Amine-functionalized polymers have the potential to encapsulate siRNA via electrostatic interaction between positively charged amines on the polymer backbone and negatively charged phosphate groups of the phosphodiester linkages of siRNA. Two parameters were investigated for siRNA loading: the impact of the PEG length and the nature of amino groups. Two polymers with different PEG lengths were investigated: 49 repeating units (long PEG), since a large body of evidence points toward a PEG length of 2,000 g/mol as the optimal length for drug delivery purposes (Jokerst et al., 2011; Knop et al., 2010; Mishra et al., 2016) and 17 units (short PEG), to determine if lowering the PEG length could promote higher siRNA loading. Primary amines were also compared to tertiary amines on the long PEG copolymer. The results are summarized in Figure 1.

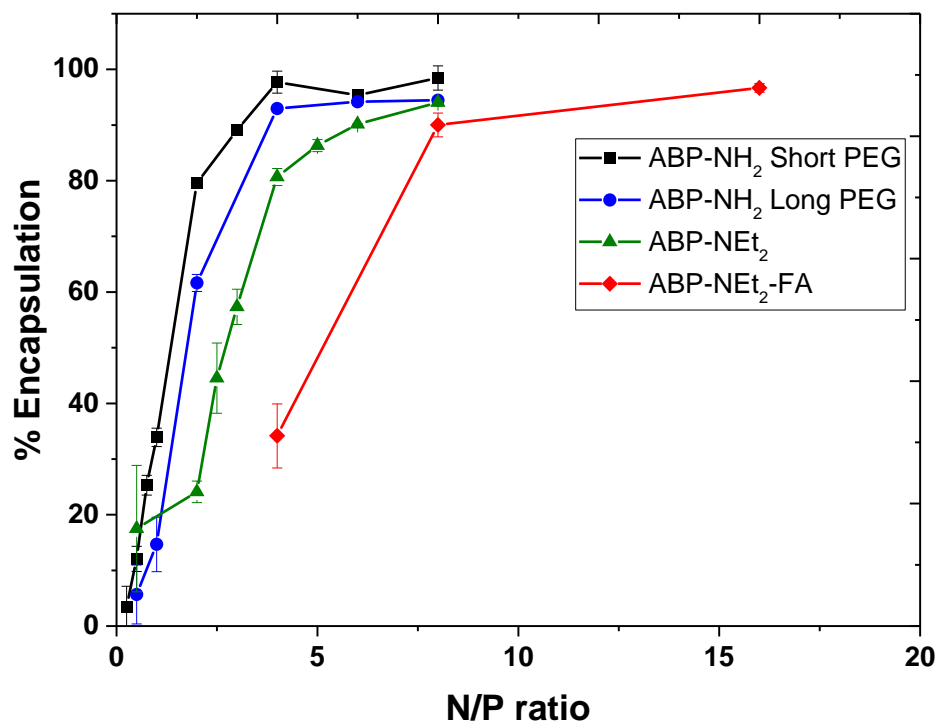

**Figure 1.** siRNA loading of CA-(PAGE-NH$_2$-*b*-PEG)$_4$ (ABP-NH$_2$) with short PEG, ABP-NH$_2$ with long PEG, CA-(PAGE-NEt$_2$-*b*-PEG)$_4$ (ABP-NEt$_2$), and CA-(PAGE-NEt$_2$-*b*-PEG-FA)$_4$ (ABP-NEt$_2$-FA). Samples were loaded with 60 nM siRNA and % encapsulation was calculated using SYBR Gold fluorescence assay (n = 3). N/P ratio is the ratio of amine over phosphate groups. At a ratio of 1.3, there are equimolar amounts of siRNA and polymers.

The results show that a longer PEG chain (49 units) was less efficient than a shorter PEG (17 units) at complexing siRNA (61 ± 2 % vs. 79 ± 1 % encapsulation at a N/P = 2, for ABP-NH$_2$). Nevertheless, both polymers were able to fully complex siRNA at N/P = 4. Only the formulation with longer PEG (CA-(AGE$_8$-*b*-EG$_{49}$)$_4$) was further characterized for gene delivery since this PEG chain length is optimal for drug delivery. Keeping the longer PEG chains constant, polymers bearing primary amines were compared to polymers functionalized with tertiary amines. CA-(PAGE-NEt$_2$-*b*-PEG)$_4$ (ABP-NEt$_2$) tertiary amine polymers demonstrated a lower ability to complex siRNA than ABP-NH$_2$, primary amine functionalized polymers, with long PEG at the same N/P ratio. Tertiary amine have a lower pK$_a$ and a lower surface potential compared to primary amines, which leads to a lower siRNA complexation strength (Lu et al., 2016; Whitehead et al., 2014). The maximum siRNA encapsulation is reached only at N/P = 8 with 94 % loading for ABP-NEt$_2$ compared to N/P = 4 for ABP-NH$_2$. Coupling of FA to the end of the PEG chains impaired siRNA association, probably because the carboxylic groups of FA are negatively charged at physiological pH (Scheme 1). The highest encapsulation efficiency was attained at a N/P of 16 with ≈ 96 % siRNA loading. As a negative control, unfunctionalized polymers (CA-(AGE$_8$-*b*-EG$_{49}$)$_4$) loaded with siRNA did not show any encapsulation efficicency of siRNA (Figure S4) corroborating the electrostatic interaction as the driving force for the siRNA loading.

### 3.3 Nanoparticle size and zetapotential

Blank and siRNA-loaded formulations were characterized by dynamic light scattering (DLS) and ζ-potential (Table 2). Interestingly, the empty ABP-NH$_2$ with long PEG formulation did not show aggregation, whereas complexation with siRNA led to the formation of aggregates of 181 nm.

**Table 2. DLS and zetapotential of empty and siRNA-loaded formulations in 5% Dextrose.**

| Formulation | N/P | Diameter (nm) | Micelle dispersity | ζ-potential (mV) |
| --- | --- | --- | --- | --- |
| ABP-NH$_2$ (long PEG) | | N/A | N/A | 0.7 ± 0.07 |
| ABP-NH$_2$ with siRNA | 4 | 181 ± 44 | 0.10 | 18.1 ± 2.5 |
| ABP-NEt$_2$ | | 209 ± 49 | 0.24 | 15.2 ± 3.9 |
| ABP-NEt$_2$ with siRNA | 16 | 188 ± 57 | 0.35 | 10.4 ± 1.0 |
| ABP-NEt$_2$-FA | | 239 ± 52 | 0.04 | 39.8 ± 2.0 |
| ABP-NEt$_2$-FA with siRNA | 16 | 188 ± 57 | 0.35 | 23.2 ± 1.7 |

Formulations were prepared at N/P ratio showing highest siRNA loading. Measurements were done in triplicates.

In the case of the tertiary amine, both ABP-NEt$_2$ and ABP-NEt$_2$-FA showed a slight decrease in size upon siRNA loading corresponding to a collapse of the nanoparticles around the siRNA structure, an observation previously reported (Katas and Alpar, 2006). Upon siRNA complexation, the polydispersity of the micelles reached 0.1 for ABP-NH$_2$ and increased from 0.24 to 0.35 for ABP-NEt$_2$ and 0.04 to 0.35 for ABP-NEt$_2$-FA. This increase in the polydispersity can result from a restructuring of the polymers around the siRNA as exemplified by the ABP-NH$_2$ formulation which do not form aggregates without siRNA but aggregates in the presence of siRNA. The ABP-NH$_2$ formulation did not show an appreciable change in zetapotential upon loading of siRNA with a variation within its standard deviation. However, both ABP-NEt$_2$ and ABP-NEt$_2$-FA showed a decrease in their ζ-potential upon siRNA loading, which is expected due to the negative charge of siRNA. Similar hydrodynamic diameters and ζ-potentials were recorded for the siRNA-loaded aggregates and allows for their comparison in cell uptake and internalization. The positive surface charge of the formulations is promising for cell uptake and transfection.

### 3.4 siRNA transfection in HeLa cells

siRNA transfection was assessed in HeLa-GFP cells to compare the primary amine polymers (ABP-NH$_2$) to the tertiary amine counterpart (ABP-NEt$_2$) (Figure 2). Unfortunately, only a small decrease in GFP expression was observed at all N/P ratios tested. It has to be noted that ABP-NH$_2$ exhibited cell toxicity at N/P > 2, whereas the lower N/P of 2 did not elicit a knockdown of GFP expression. Although less toxic, the ABP-NEt$_2$ did not yield substantial GFP knockdown at all N/P ratio tested. Several strategies were used to improve transfection, by increasing the siRNA concentration up to 60 nM, by removing serum in the culture media and by adding chloroquine, but none of them improved the transfection efficiency of ABP-NEt$_2$, whereas commercial Lipofectamine was able to silence GFP to the extent of 51.2 % (Figure S7).

To further investigate the limited transfection of the ABP-NEt$_2$ formulation, cell uptake was quantified in HeLa model cells using fluorescently labeled siRNA (Figure 2). The results show a low cell uptake after 2, 4, and 24 h incubation time with less than 5 % uptake even at the longer time points. To promote higher cell uptake, folic acid (FA) was conjugated to the end of the PEG chains of ABP-NEt$_2$ yielding ABP-NEt$_2$-FA. Unexpectedly, this approach did not improve the cell uptake in HeLa cells, since less than 5 % of the formulation was internalized in the cell after 24 h, whereas Lipofectamine was able to enter the cells (Figure 2).

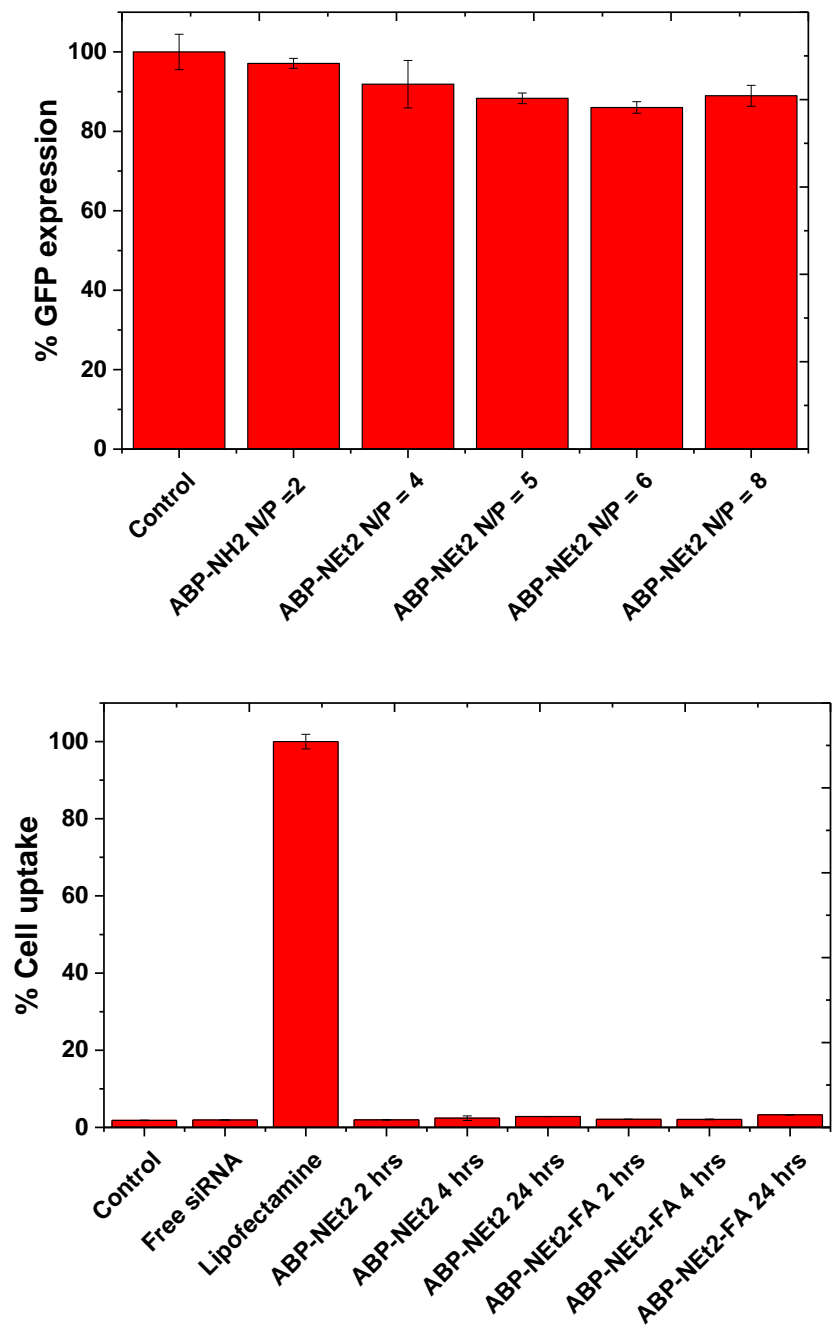

**Figure 2.** (top) siRNA transfection of ABP-NH$_2$ and ABP-NEt$_2$ in HeLa-GFP measured by flow cytometry of GFP fluorescence after 48 hrs incubation (n = 3). siRNA concentration was maintained at 30 nM and the N/P ratio varied. (bottom) Cell uptake of ABP-NEt$_2$ and ABP-NEt$_2$-

FA in HeLa cells assessed by flow cytometry. 60 nM of Alexa488-labeled siRNA was complexed to the NP with a N/P ratio of 16. The results were obtained as triplicates.

To corroborate these results, the internalization was qualitatively monitored by fluorescence microscopy using Cy5-siRNA-loaded formulations (Figure 3). Lipofectamine used as a positive control show a large uptake with the cells appearing bright red, whereas the three formulations ABP-NH$_2$, ABP-NEt$_2$ and ABP-NEt$_2$-FA did not show an appreciable uptake. Only ABP-NEt$_2$-FA formulation showed a small uptake with a faint red appearing inside the cells. Interestingly, aggregates of ≈ 1-2 µm appearing as bright red spots were seen in the microscopy images of all three formulations and were absent in the lipofectamine positive control. These aggregates could indicate a colloidal instability of the formulations in the cell culture media. To better understand this colloidal instability, the variation of the size of the ABP-NEt$_2$-FA nanoparticles were measured by DLS in different solutions as a function of temperature (Figure S8). The results show that in 5% Dextrose, ABP-NEt$_2$-FA do not show a LCST behavior, whereas in DMEM cell culture media, ABP-NEt$_2$-FA exhibits a LCST value of 37 °C. The presence of the LCST can be attributed to a salting-out effect on the polymer due to the high salt concentrations in DMEM (Du et al., 2010; Freitag and Garret-Flaudy, 2002; Liu et al., 2012). In order to optimize the colloidal stability of ABP-NEt$_2$-FA, the addition of lipids as co-surfactants was investigated. DOPE, DSPE-PEG$_{2k}$, and DSPE-PEG$_{5k}$ were added at a concentration of 10 wt%. The addition of the lipids shifted the LCST in DMEM to 42 °C for DOPE and DSPE-PEG$_{5k}$ and 44 °C for DSPE-PEG$_{2k}$. Adding lipids as co-surfactant has the potential to provide colloidal stability of the ABP-NEt$_2$-FA formulation and optimize the siRNA delivery (Harmon et al., 2011; Le Devedec et al., 2013; Li et al., 2011b; Wu et al., 2013).

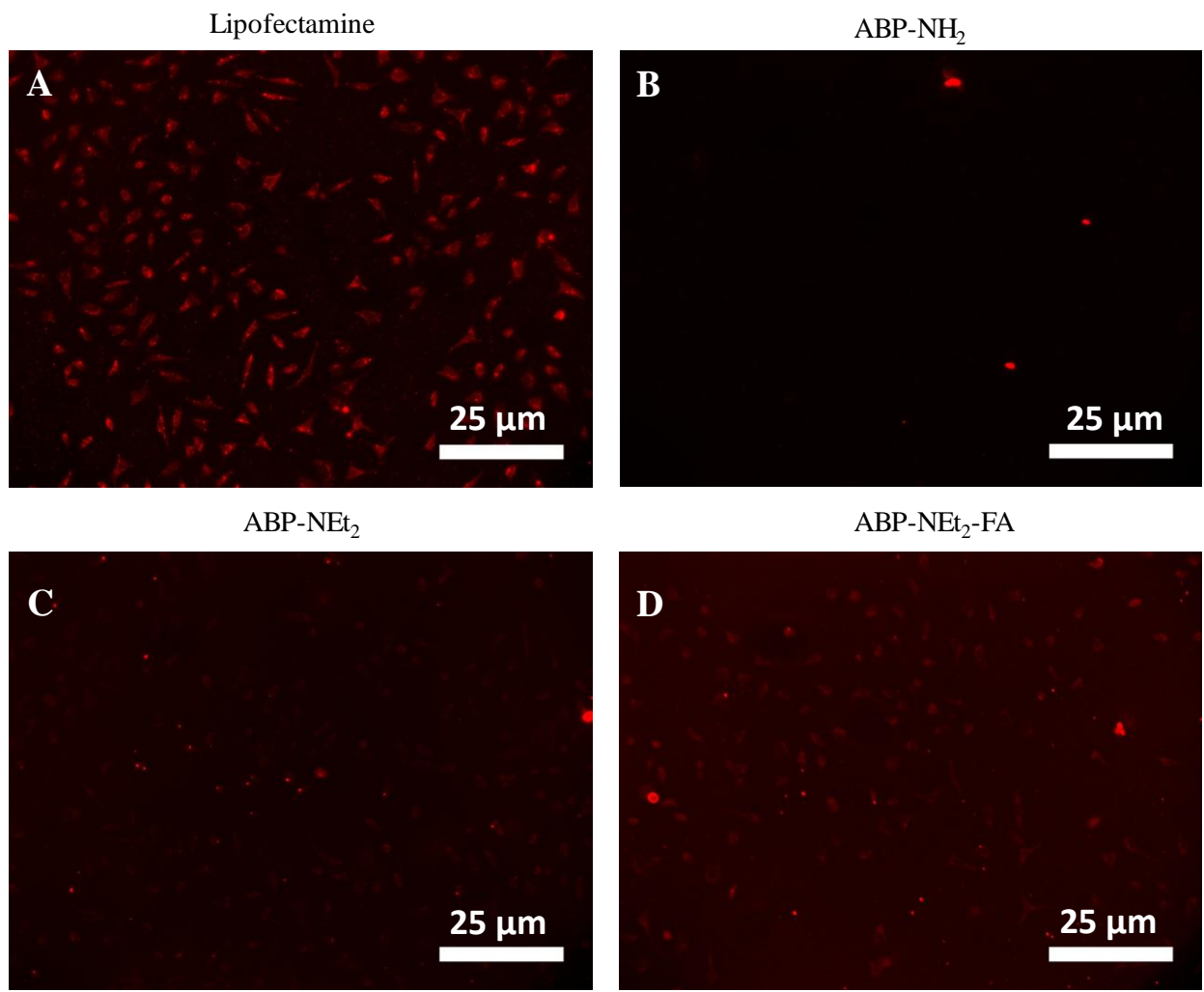

**Figure 3.** Fluorescence microscopy images for cell uptake of (A) lipofectamine used as a control and (B) ABP-NH$_2$, (C) ABP-NEt$_2$ and (D) ABP-NEt$_2$-FA in HeLa cells. 100 nM of Cy5-labeled siRNA was complexed to the NP with a N/P ratio of 16 for 30 min. The results were imaged at a 10X magnification and the same brightness level was applied to each image. The scale bars represent 25 μm in length.

### 3.5 siRNA encapsulation using lipid-polymer mixed micelles (LPM)

Since the ABP-NEt$_2$-FA yielded better cell uptake compared to the other formulations (Figure 3), this formulation was further studied for stabilisation with co-lipids. A series of three lipids were chosen to form LPMs namely DOPE, DSPE-PEG$_{2k}$ and DSPE-PEG$_{5k}$. DOPE was chosen as it has been often used to improve transfection efficiency (Wasungu and Hoekstra, 2006). Also, DSPE lipids were tested as co-surfactants since they are used in the FDA-approved drug delivery system Doxil® (Soundararajan et al., 2009). To ascertain the formation of mixed micelles rather than separate aggregates of the polymer and lipids, the hydrodynamic diameter of the ABP-NEt$_2$-FA was measured in the absence and presence of the lipids (Table 3). DSPE-PEG$_{2k}$ and DSPE-PEG$_{5k}$ alone at 0.1 mg/mL failed to form stable aggregates in 5% Dextrose using the coevaporation method (data not shown). Whereas, DOPE at a concentration of 0.1 mg/mL forms micelles with diameters of 188 ± 76 nm with a micelle dispersity of 0.19 (Figure S9). To form the mixed micelles, ABP-NEt$_2$-FA was mixed with 10 wt% lipids using the coevaporation method (Ciani et al., 2004; Harmon et al., 2011). The mixed micelles showed a large increase in the micelle dispersity since ABP-NEt$_2$-FA micelles had a dispersity of 0.04 compared to 0.28 - 0.54 for mixed micelles. Interestingly, the size of the ABP-NEt$_2$-FA NP decreased substantially upon addition of the lipids to form the LPMs, an observation that was previously reported (Li et al., 2011a). The surface ζ-potential of ABP-NEt$_2$-FA with DOPE was higher than ABP-NEt$_2$-FA alone, 51.6 mV compared to 39.8 mV, due to the presence of positive charges on DOPE lipid. On the contrary, adding PEGylated lipids to the formulation lowered the surface zetapotential of ABP-NEt$_2$-FA (Table 3). The presence of the PEG chains masks the surface ζ-potential and a trend is observed where the longer the PEG chains are, the lower the ζ-potential (Kumar et al., 2014). To further corroborate the formation of mixed micelles, DSC thermograms of the mixed micelles were compared to ABP-NEt$_2$-FA, DOPE, DSPE-PEG$_{2k}$, and DSPE-PEG$_{5k}$ alone (Figure S10-S12). ABP-NEt$_2$-FA showed a melting temperature of 43.1 °C (Figure S10). The lipids alone have a melting point of 115.4 °C for DOPE, 50.3 °C for DSPE-PEG$_{2k}$, and 57.7 °C for DSPE-PEG$_{5k}$ (Figure S11). When lipids are incorporated in the polymers to form mixed micelles, the melting points of the lipids disappeared (Figure S12). Also, the DSC shows the appearance of a glass transition temperature at 25 °C for the ABP-NEt$_2$-FA with DOPE, 12 °C for ABP-NEt$_2$-FA with DSPE-PEG$_{2k}$, and 16 °C for ABP-NEt$_2$-FA with DSPE-PEG$_{5k}$. In the thermograms of the mixed micelles, only a subtle peak of a melting point can be observed between 45.7 and 51.4 °C which should correspond to the ABP-NEt$_2$-FA polymer chains that are free from interactions.

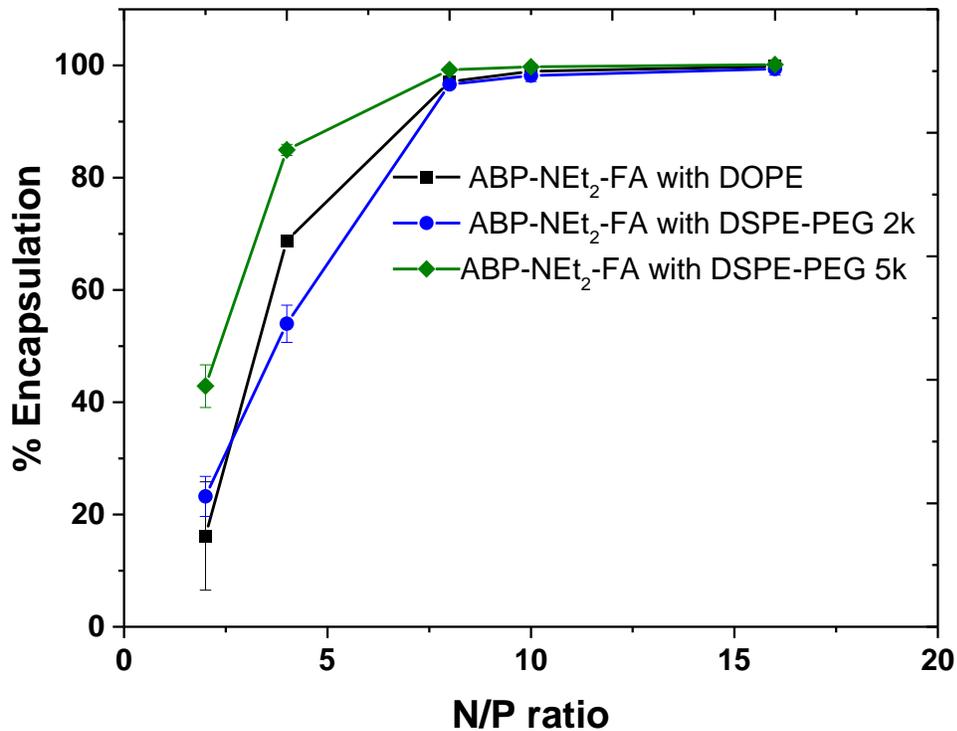

**Figure 4.** siRNA loading of ABP-NEt$_2$-FA with DOPE, DSPE-PEG$_{2k}$ and DSPE-PEG$_{5k}$. Samples were loaded with 60 nM siRNA and % encapsulation was calculated against a standard curve. Samples were obtained in triplicates.

As shown in Figure 4, the presence of lipids improved the loading of siRNA for all three formulations compared to ABP-NEt$_2$-FA alone but yielded lower encapsulation capacity than the ABP-NEt$_2$ and ABP-NH$_2$ formulations. Nevertheless, all three formulations show close to 100% encapsulation at N/P of 8, similarly to ABP-NEt$_2$-FA. Size and $\zeta$-potential of the mixed micelles formulation before and after siRNA loading are presented in Table 3.

**Table 3. DLS and zetapotential of empty and siRNA-loaded formulations in 5% Dextrose.**

| siRNA formulation | Diameter (nm) | Micelle dispersity | $\zeta$-potential (mV) |
| --- | --- | --- | --- |
| ABP-NEt$_2$-FA | 239 ± 52 | 0.04 | 39.8 ± 2.0 |
| ABP-NEt$_2$-FA with siRNA | 188 ± 57 | 0.35 | 23.2 ± 1.7 |
| ABP-NEt$_2$-FA/DOPE | 103 ± 55 | 0.28 | 51.6 ± 2.5 |
| ABP-NEt$_2$-FA/DOPE with siRNA | 221 ± 54 | 0.01 | 30.9 ± 2.6 |
| ABP-NEt$_2$-FA/DSPE-PEG$_{2k}$ | 139 ± 62 | 0.38 | 32.9 ± 3.8 |
| ABP-NEt$_2$-FA/DSPE-PEG$_{2k}$ with siRNA | 206 ± 64 | 0.13 | 19.5 ± 0.2 |
| ABP-NEt$_2$-FA/DSPE-PEG$_{5k}$ | 121 ± 55 | 0.54 | 26.8 ± 1.5 |
| ABP-NEt$_2$-FA/DSPE-PEG$_{5k}$ with siRNA | 153 ± 49 | 0.24 | 14.9 ± 1.6 |

Formulations were prepared at N/P ratio of 16. Measurements were done in triplicates.

Upon siRNA loading there is an increase in the micelle size with a concomitant decrease in the micelle dispersity. However, the micelle size of the siRNA-loaded mixed micelles are comparable to the ABP-NEt$_2$-FA with siRNA. Also, upon siRNA loading there is a decrease of the zetapotential, an observation that has been previously reported (Abdul Ghafoor Raja et al., 2015). To ensure that the polymer and polymer-lipid mixed micelles were non toxic and could be used as siRNA delivery systems, the cytotoxicity of each polymer as well as the blank micelles was measured in HeLa cells using the MTS assay (Figure S13). ABP-NH$_2$ was quite toxic as soon as 0.05 mg/mL and decreased cell viability to 32% at 1 mg/mL, which could explain the poor results of transfection (Figure 2). ABP-NEt$_2$ was less noxious to the cells, but followed the same trend, since only 48% of the cells were viable at 1 mg/mL. Interestingly, ABP-NEt$_2$-FA was much better tolerated up to 1 mg/mL. Although some reports show no significant difference in cytotoxicity upon FA conjugation (Arote et al., 2010; Cheng et al., 2009), in this report the FA improved the biocompatibility of the system; an observation that has been previously reported (Liu et al., 2012). It is hypothesized that the folate ligand shields partially the interaction between the positive charges of the polymer and cell surface. Of the polymer-lipid mixed micelles, ABP-NEt$_2$-FA with DSPE-PEG$_{2k}$ or DSPE-PEG$_{5k}$ were well tolerated with close to 100% cell viability up to 1 mg/mL, whereas ABP-NEt$_2$-FA with DOPE showed higher toxicity with 72% cell viability at 0.5 mg/mL.

### 3.6 siRNA transfection with lipid-polymer mixed micelles (LPM)

To determine the potential of the mixed micelles towards siRNA transfection, the cell uptake was observed using fluorescence microscopy (Figure 5). The results show a marked increase of the mixed micelles uptake in HeLa cells when compared to the micelles without lipids as cosurfactants. ABP-NEt$_2$-FA with DSPE-PEG$_{2k}$ and ABP-NEt$_2$-FA with DSPE-PEG$_{5k}$ both show higher uptake than the ABP-NEt$_2$-FA with DOPE. Moreover, the large aggregates seen in the previous formulation are absent in these images.

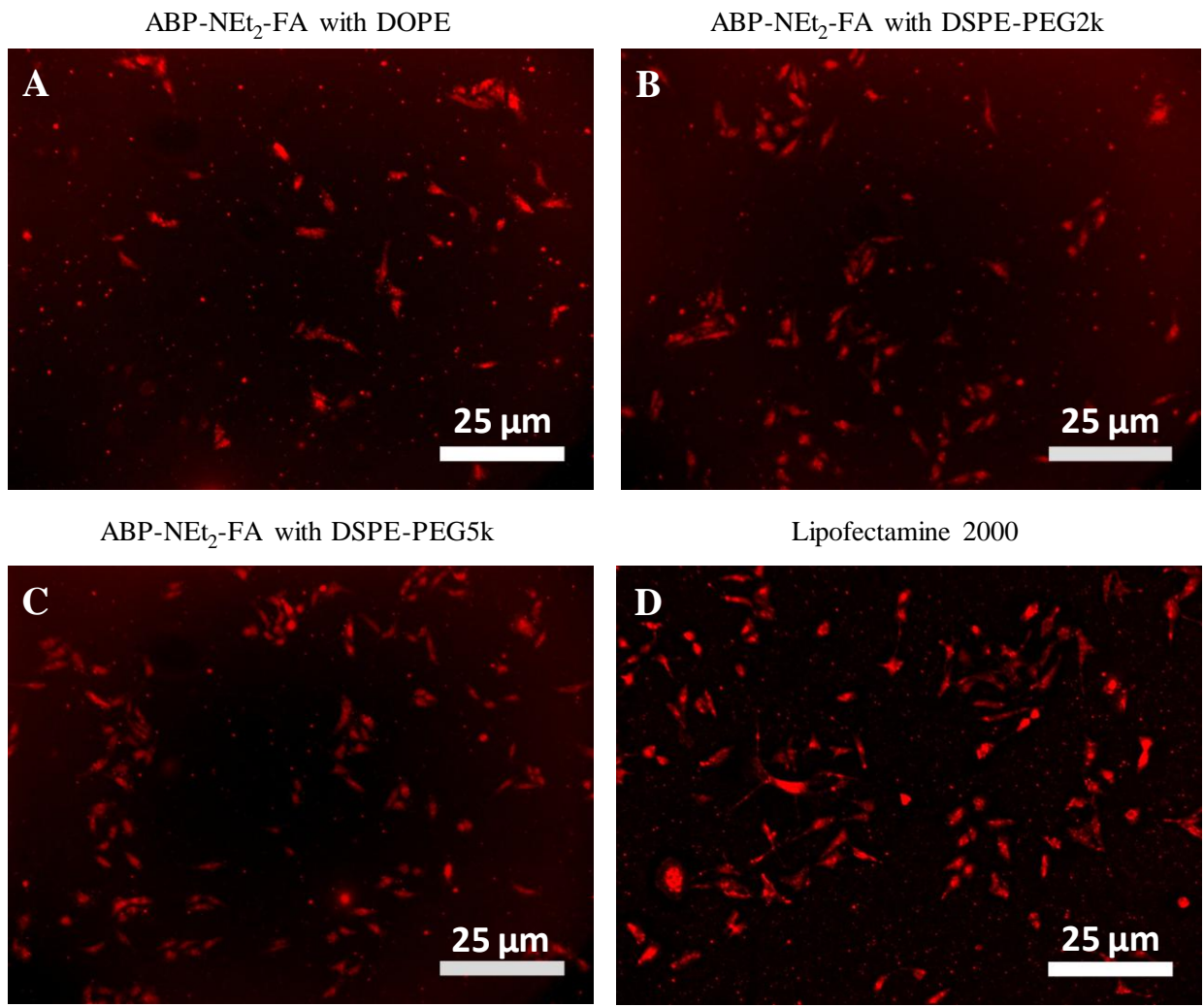

**Figure 5.** Fluorescence microscopy images for cell uptake (A) ABP-NEt$_2$-FA with DOPE, (B) ABP-NEt$_2$-FA with DSPE-PEG$_{2k}$, (C) ABP-NEt$_2$-FA with DSPE-PEG$_{5k}$ and (D) Lipofectamine 2000 in HeLa cells. 100 nM of Cy5-labeled siRNA was complexed to the NP with a N/P ratio of 16 for 30 min. The results were imaged at a 10X magnification and the same brightness level was applied to each image. The scale bars represent 25 μm in length.

These qualitative observations were confirmed by flow cytometry (Figure 6). When normalized to lipofectamine, mixed micelles exhibited 18-24% higher uptake than polymeric micelles, but still 5 times lower than lipofectamine. The presence of lipids likely aided in the stabilization of the nanoparticles in the culture medium by increasing the LCST thereby preventing aggregation, enabling cellular uptake. As compared to non-targeted micelles, the presence of FA significantly improved the uptake of the micelles (from ≈ 5% to 18-24%). This demonstrated the advantage of both components: lipids for improved stability and FA-conjugation for improved cell uptake.

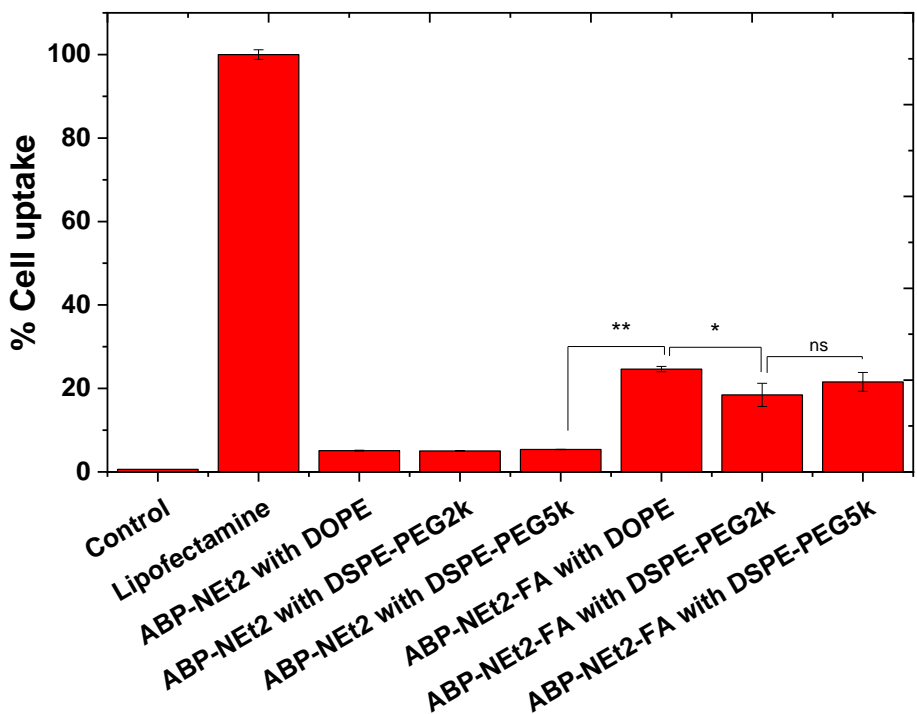

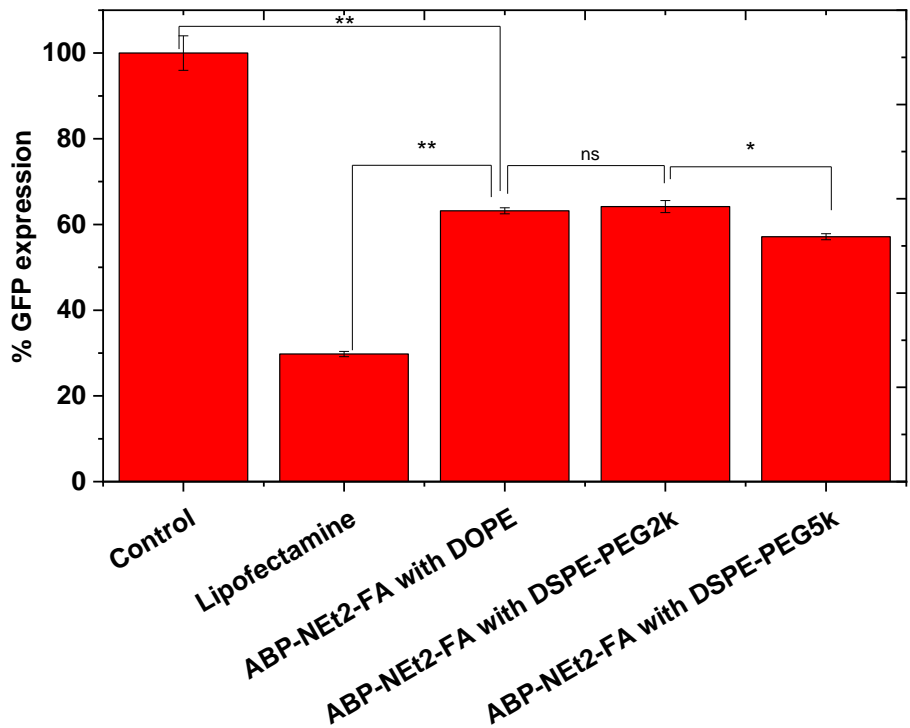

**Figure 6.** (top) Cell uptake of mixed micelles in HeLa cells assessed by flow cytometry. 60 nM of Alexa488-labeled siRNA was complexed to the NPs with a N/P ratio of 16. The results were obtained as triplicates. (bottom) siRNA transfection of mixed micelles in HeLa-GFP measured by flow cytometry of GFP fluorescence after 48 hrs incubation (n = 3). siRNA concentration was maintained at 60 nM and N/P of 16. Statistical analysis performed with Student's t test * = $p < 0.05$; ** = $p < 0.01$; ns = $p > 0.05$.

Although the uptake level did not reach Lipofectamine performance, this increase was enough to trigger significant siRNA transfection (Figure 6). ABP-NEt$_2$-FA mixed micelles reduced GFP expression to 57-65%, showing similar behaviour for all three lipids, DOPE, DSPE-PEG$_{2k}$ and DSPE-PEG$_{5k}$. Although not as efficient as lipofectamine, these results testify to an improvement in the formulation with the presence of the lipids as cosurfactant. The literature demonstrates the beneficial effect of DOPE in lipid-based delivery systems due to its fusogenic effect (Du et al., 2014; Gjetting et al., 2010; Mochizuki et al., 2013). To determine the extent of lysosomal escape of the LPM, siRNA-loaded LPM and lysosomes colocalization was observed under fluorescence microscopy (Figure S14). Both LysoTracker® green and siRNA labeled with Cy 5.5 were readily taken up by the cells after 4 h. For all three formulations of ABP-NEt$_2$-FA mixed micelles, strong colocalization was observed, confirming that internalization of LMP followed the endocytotic pathway. This observation suggested that endosomal entrapment decreased the transfection efficiency and might explain the lower transfection observed for the mixed micelles when compared to lipofectamine (Figure 6).

The proposed BA-based block copolymers were developed for drug and gene delivery due to their highly biocompatible profile. In this study, the star-shaped BA-based block copolymers show a high siRNA complexation at low N/P ratio of 16 which is advantageous for gene delivery purposes because limited material is needed to deliver the siRNA, which decreases the potential toxic effects (Hall et al., 2017). Adding co-lipids improved the loading and enabled transfection of the star-shaped block copolymers. An improvement to the proposed drug delivery system would be to optimize the choice of the lipid co-surfactant in order to increase the silencing activity to reach comparable levels to lipofectamine. In this perspective, pH-switchable lipids, which demonstrated quick endosomal escape and efficient cytoplasmic release of siRNA, could be used (Viricel et al., 2017). An advantage of the LPM system is the ease with which the platform can be tuned. The formulation can be optimized through the selection of lipids without the need to change the entire polymer structure. However, further studies are necessary to determine the stability of the polymer-lipid mixed micelles in the presence of serum protein. Lipid-polymer hybrid nanoparticles are gaining importance because these systems combine the advantages of both polymers and lipids (Hadinoto et al., 2013). In one study, lipid-polymer hybrids enabled sustained long-term release of siRNA from nanoparticles which is an improvement to the transient release typically seen in nanoparticles made solely of lipids or polymers (Shi et al., 2014). Another report used PEG-*b*-PLA-*b*-PEG polymer and DDAB lipid hybrid nanoparticles to improve the colloidal stability of DDAB in RPMI and promote gene silencing (monirinasab et al., 2018). The advantage of using polymer-lipid hybrid nanoparticles in our system is that the polymer part confers lower cytotoxicity (Figure S13) and high siRNA loading, whereas the lipid part promotes stability in the cell culture medium and high transfection.

## 5. Conclusion

Cholic acid-based block copolymers were obtained by anionic polymerization. The PAGE blocks may be functionalized to bear pendant primary or tertiary amine groups enabling siRNA loading. High loading was obtained for both primary and tertiary amines with close to 100% complexation at low N/P ratio. Folic acid can be grafted to the ends of the PEG chains for increased cell uptake in cancer cells. Unfortunately, the siRNA loaded micelles did not exhibit cell uptake or siRNA transfection. DLS temperature trends demonstrated that the nanoparticles were unstable in the DMEM cell culture media due to the presence of a LCST at 37 °C. DOPE, DSPE-PEG$_{2k}$, and DSPE-PEG$_{5k}$ were used as co-surfactants to increase the colloidal stability of the ABP-NEt$_2$-FA NP. Adding these co-surfactants allowed to increase the LCST to 42 °C for DOPE and DSPE-PEG$_{5k}$ and 44 °C for DSPE-PEG$_{2k}$. Although not as efficient, the presence of the lipids did not inhibit siRNA loading with close to 100% complexation at N/P ratio of 16. DSC results testify to the presence of mixed micelles with the disappearance of the melting endotherms of both ABP-NEt$_2$-FA polymers and corresponding lipids and the appearance of a glass transition temperatures for the mixed micelles. Finally, adding the mixed micelles showed an improved cell uptake. Although, not as efficient as lipofectamine, the mixed micelles showed siRNA knockdown of GFP expression. These results show that there is a great potential towards the use of cholic acid-based block copolymers for the delivery of siRNA in the form of mixed micelles. The star-shaped architecture of the cholic acid-based polymers enabled high siRNA loading but the presence of high salt concentrations in the cell culture media led to colloidal instability, low cell uptake and low siRNA transfection. Fortunately, the presence of lipids as co-surfactants allowed to increase the LCST of the mixed micelles and provided colloidal stability in cell culture medium and promote siRNA transfection. Adding lipids to the polymer nanoparticles offers new opportunities in formulations of drug delivery systems. There are a large variety of modifications that can be added to improve polymer-based systems, such as adding targeting lipids to improve cell uptake, lipids as co-surfactants to improve colloidal stability, fusogenic lipids to increase endosomal release and even pH-sensitive lipids to promote drug release.

The financial support from NSERC of Canada in the form of Discovery grant to XXZ is gratefully acknowledged. AC and XXZ are members of CQMF funded by FRQNT of Quebec.